\documentclass[11pt,a4paper,hidelinks]{article}
\usepackage[utf8]{inputenc}
\usepackage[english]{babel}
\usepackage{amsmath}
\usepackage{amsfonts}
\usepackage{amssymb}
\usepackage{graphicx}
\usepackage{authblk}
\usepackage{subcaption}
\usepackage{cite}
\usepackage{booktabs}
\usepackage{multirow}
\usepackage{bm}
\usepackage{nicefrac}
\usepackage{xcolor}
\usepackage{microtype}
\usepackage{setspace}

\usepackage[left=3cm,right=3cm,top=2cm,bottom=2cm]{geometry}

\usepackage{array}

\usepackage{setspace}

\title{Context encoding enables machine learning based quantitative photoacoustics}
\date{June 12th, 2017}
\author[1,2]{\underline{Thomas Kirchner}\,\thanks{Please address all correspondence to t.kirchner@dkfz-heidelberg.de or l.maier-hein@dkfz-heidelberg.de}}
\author[1,3]{\underline{Janek Gröhl}\,}
\author[1,3]{Lena Maier-Hein\,$^*$}
\affil[1]{Division of Computer Assisted Medical Interventions (CAMI), German Cancer Research Center (DKFZ), Heidelberg, Germany}
\affil[2]{Faculty of Physics and Astronomy, Heidelberg University, Germany}
\affil[3]{Medical Faculty, Heidelberg University, Germany}
\begin{document}

\maketitle
%\thanks{}

\begin{abstract}
Real-time monitoring of functional tissue parameters, such as local blood oxygenation, based on optical imaging could provide groundbreaking advances in the diagnosis and interventional therapy of various diseases. While photoacoustic (PA) imaging is a novel modality with great potential to measure optical absorption deep inside tissue, quantification of the measurements remains a major challenge. In this paper, we introduce the first machine learning based approach to quantitative PA imaging (qPAI), which relies on learning the fluence in a voxel to deduce the corresponding optical absorption. The method encodes relevant information of the measured signal and the characteristics of the imaging system in voxel-based feature vectors, which allow the generation of thousands of training samples from a single simulated PA image. Comprehensive \emph{in silico} experiments suggest that context encoding (CE)-qPAI enables highly accurate and robust quantification of the local fluence and thereby the optical absorption from PA images.
\end{abstract}


\section*{Introduction}
Photoacoustic (PA) imaging is a novel imaging concept with a high potential for real-time monitoring of functional tissue parameters such as blood oxygenation deep inside tissue. It measures the acoustic waves arising from the stress-confined thermal response of optical absorption in tissue \cite{Wang2016-nb}. More specifically, a photoacoustic signal $S(\bm{v})$ in a location $\bm{v}$ is a pressure response to the locally absorbed energy $H(\bm{v})$, which, in turn, is a product of the absorption coefficient $\mu_\textrm{a}(\bm{v})$, the Grueneisen coefficient $\Gamma(\bm{v})$ and the light fluence $\phi(\bm{v})$.
\begin{equation}
    S(\bm{v}) \propto H(\bm{v}) = \mu_\textrm{a}(\bm{v}) \cdot \Gamma(\bm{v}) \cdot \phi(\bm{v})
    \label{eqSignal}
\end{equation}
Given that the local light fluence not only depends on the imaging setup but is also highly dependent on the optical properties of the surrounding tissue, quantification of optical absorption based on the measured PA signal is a major challenge \cite{Wang2012-ms,Cox2009-vn}.
So far, the field of quantitative PA imaging (qPAI) has focussed on model-based iterative optimization approaches to infer optical tissue parameters from measured signals (cf. e.g. \cite{Cox2009-vn,Iftimia2000-bb,Cox2005-tl,Cox2006-hw,Yuan2006-qx,Laufer2007-va,Malone2016-ew,Haltmeier2015-eq,Cox2012-ao,Banerjee2008-rj}). While these methods are well-suited for tomographic devices with high image quality (cf. e.g. \cite{Wang2009-op,Xia2014-hs,Tzoumas2016-yt}) as used in small animal imaging, translational PA research with clinical ultrasound transducers or similar handheld devices (cf. e.g. \cite{Niederhauser2005-wk,Zackrisson2014-pb,Wang2016-nb,Upputuri2017-fz,Gamelin2008-we,Song2008-rr,Kim2010-cu,Garcia-Uribe2015-oi}) focusses on qualitative image analysis.\par

As an initial step towards clinical qPAI, we introduce a novel machine learning based approach to quantifying PA measurements. The approach features high robustness to noise while being computationally efficient. In contrast to all other approaches proposed to date, our method relies on learning the light fluence on a voxel level to deduce the corresponding optical absorption. Our core contribution is the development of a voxel-based \emph{context image} (CI) that encodes relevant information of the measured signal voxel together with characteristics of the imaging system in a single feature vector. This enables us to tackle the challenge of fluence estimation as a machine learning problem that we can solve in a fast and robust manner. Comprehensive \emph{in silico} experiments indicate high accuracy, speed, and robustness of the proposed context encoding (CE)-qPAI approach. This is demonstrated for estimation of (1) fluence and optical absorption from PA images, as well as (2)  blood oxygen saturation as an example of functional imaging using multispectral PA images.\par

\section*{Materials and Methods}
A common challenge when applying machine learning methods to biomedical imaging problems is the lack of labeled training data. In the context of PAI, a major issue is the strong dependence of the signal on the surrounding tissue. This renders separation of voxels from their context - as in surface optical imaging \cite{Wirkert2016-yz} - impossible or highly inaccurate. 
Simulation of a sufficient number of training volumes covering a large range of tissue parameter variations, on the other hand, is computationally not feasible given the generally long runtime of Monte Carlo methods which are currently the gold standard for the simulation of light transportation in tissue \cite{Cox2012-ao}.\par
\begin{figure}[htb!]
    \includegraphics{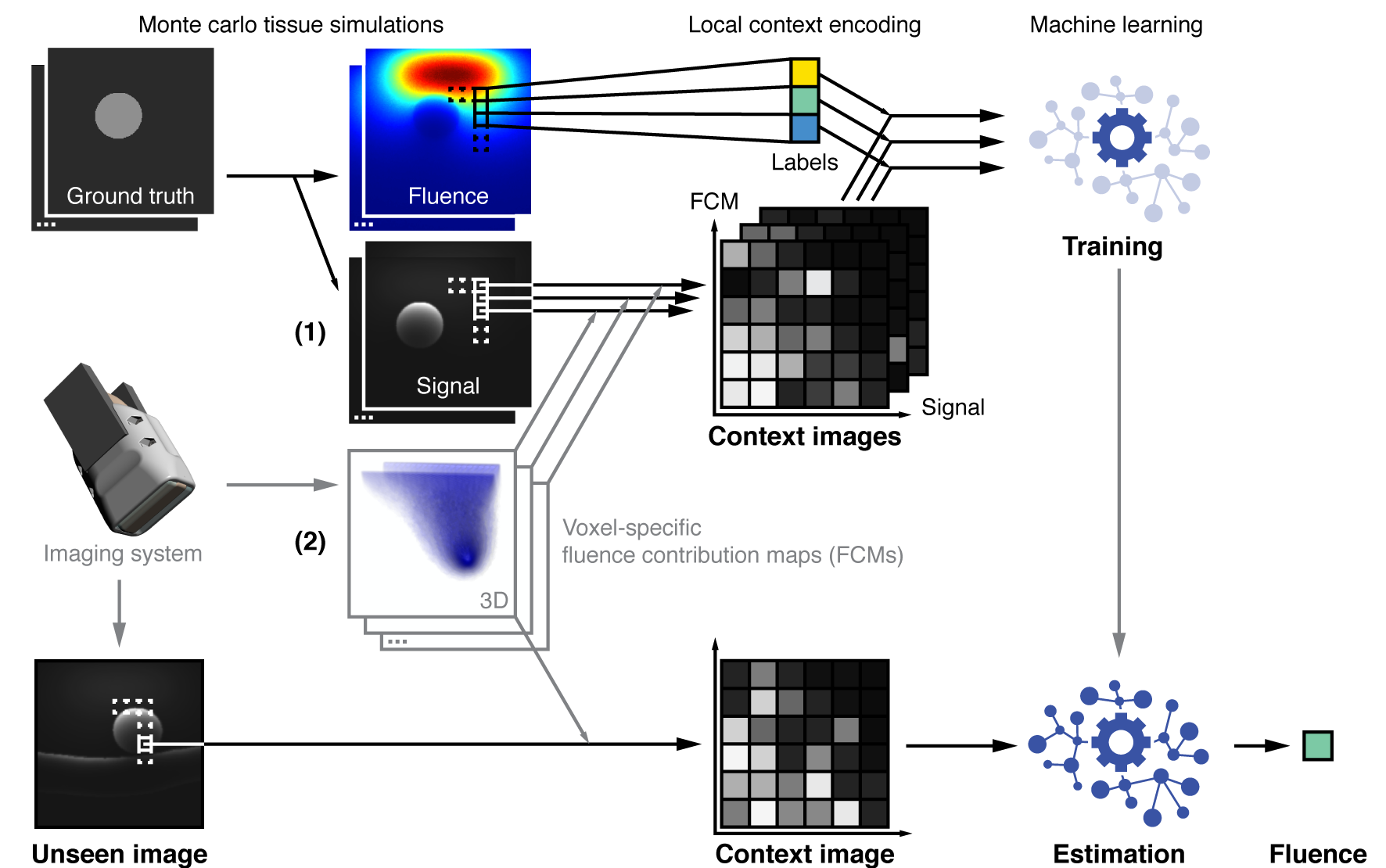}\\
    \caption[Machine learning approach to fluence estimation with context images.]{\textbf{Machine learning approach to fluence estimation with context images.} Context images (CIs) are generated individually for each voxel and encode both (1) relevant information on the measured signal extracted from the PAI signal volume  and (2) prior knowledge on the characteristics of the imaging system represented by fluence contribution maps (FCMs). During algorithm training, a regressor is presented tuples of context images and corresponding ground truth fluence values for each voxel in the training data. For estimation of optical absorption in voxels of a previously unseen image, the voxel-specific context image is generated and used to infer the local fluence using the trained regressor.}
    \label{figMethod}
\end{figure}

Inspired by an approach to shape matching, where the shape context is encoded in a so-called \emph{spin image} specifically for each node in a mesh \cite{Johnson1999-ct}, we encode the voxel-specific context in so-called context images (CIs). This allows us to train machine learning algorithms on a voxel level rather than image level and thus require orders of magnitude fewer simulated training volumes. CIs encode relevant information of the measured signal as well as characteristics of the imaging system (represented by so-called voxel-specific fluence contribution maps (FCMs)). The CIs serve as a feature vector for said machine learning algorithm which are trained to estimate fluence in a voxel. The entire quantification method is illustrated in Figure\,\ref{figMethod} which serves as an overview with details given in the following subsections. \par

\subsection*{Fluence Contribution Map}
An important prerequisite for computing the CI for a voxel $\bm{v}$ is the computation of the corresponding FCM, referred to as $\textrm{FCM}[\bm{v}]$.  $\textrm{FCM}[\bm{v}](\bm{v'})$ represents a measure for the likelihood that a photon arriving in voxel $\bm{v}$ has passed $\bm{v'}$. In other words, a FCM reflects the impact of a PA signal in $\bm{v'}$ on the drop in fluence in voxel $\bm{v}$. An illustration of a FCM corresponding to a typical handheld PA setup is shown in Figure\,1\,(2). The $\textrm{FCM}[\bm{v}]$ is dependent on how the PA excitation light pulse propagates through homogeneous tissue to arrive in $\bm{v}$ given a chosen hardware setup. The $x \times y$ FCMs per imaging plane are generated once for each new hardware setup and each voxel in the imaging plane. \par 

In this first implementation of the CE-qPAI concept, FCMs are simulated with the same resolution as the input data assuming a background absorption coefficient of 0.1\,cm$^{-1}$ and a constant reduced scattering coefficient of 15\,cm$^{-1}$ \cite{Jacques2013-pm}. The number of photons is varied to achieve a consistent photon count in the target voxel. The FCMs are generated with the widely used Monte Carlo simulation tool \emph{mcxyz} \cite{Jacques2014-ia}. We integrated mcxyz into the open-source Medical Image Interaction Toolkit MITK \cite{Wolf2004-jw} as mitkMcxyz and modified it to work in a multi-threaded environment. Sample FCMs for different voxels are illustrated in Figure\,\ref{figCIandFCM}, which also shows the generation of context images.\par

\subsection*{Context Image}
\begin{figure}[tbh!]
    \includegraphics{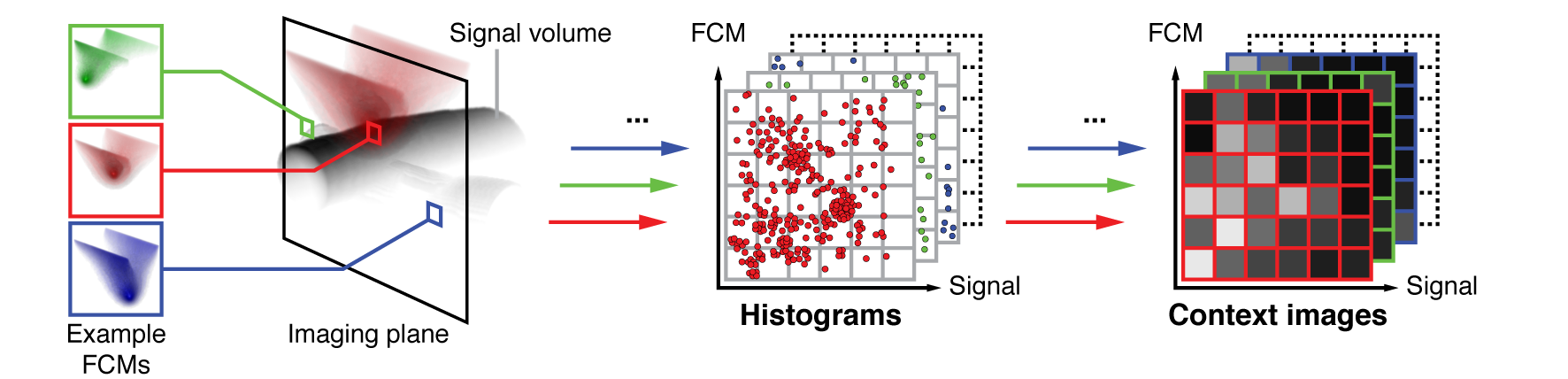}
    \caption[Context Images from Fluence Contribution Maps]{\textbf{Generation of context images} using FCMs and PA signal volumes. The voxel-specific FCMs serve as a representation of how the PA excitation light pulse propagates through homogeneous tissue to arrive in a target voxel given a chosen hardware setup. Signal volume and FCM are superimposed to generate 2D histograms from tuples of measured signal and corresponding fluence contribution.}
    \label{figCIandFCM}
\end{figure}

The CI for a voxel $\bm{v}$ in a PA volume is essentially a 2D histogram composed of (1) the measured PA signal S in the tissue surrounding $\bm{v}$ and (2) the corresponding $\textrm{FCM}[\bm{v}]$. More specifically, it is constructed from the tuples $\{(\textrm{S}(\bm{v'}), \textrm{FCM}[\bm{v}](\bm{v'})) | \bm{v'} \in N(\bm{v})\}$ where $N(\bm{v})$ is defined as $N(\bm{v}) = \{\bm{v'} | \textrm{FCM}[\bm{v}](\bm{v'}) > \epsilon\}$. This constraint is set to exclude voxels with a negligible contribution to the fluence in $\bm{v}$. The tuples are arranged by magnitude of $\textrm{S}(\bm{v'})$ and $\textrm{FCM}[\bm{v}](\bm{v'})$ into a 2D histogram and thereby encode the relevant context information in a compact form. 
In our prototype implementation of the CE-qPAI concept, the fluence contribution and signal axes of the histogram are discretized in 12 bins and scaled logarithmically to better represent the predominantly low signal and fluence contribution components. The ranges of the axes are set as $0 < \log(S) < \log(\textrm{255})$ and $\log(\epsilon) < \log(\textrm{FCM}) < -1$. Signals and fluence contributions larger than the upper boundary are included in the highest bin while smaller signals and fluence contributions are not. Figure\,\ref{figCIandFCM} illustrates the generation of CIs from FCMs and PA signals. Labelled CIs are used for training a regressor that can later estimate fluence, which, in turn, is used to reconstruct absorption (Eq.\,\ref{eqSignal}).\par
%The absorption coefficient  $\hat{\mu}_\textrm{a}(\bm{v_\textrm{u}})$ is then derived from $\hat{\mu}_\textrm{a}(\bm{v_\textrm{u}}) = S(\bm{v_\textrm{u}}) / (\hat{\phi}(\bm{v_\textrm{u}}) \cdot \Gamma(\bm{v_\textrm{u}}))$ where we assume the Grueneisen coefficient $\Gamma$ as constant over all voxels.

\subsection*{Machine learning based regression for fluence estimation}
During the training phase, a regressor is presented tuples $(\textrm{CI}(\bm{v}), \phi(\bm{v}))$ of context images $\textrm{CI}(\bm{v})$ and corresponding ground truth fluence values $\phi(\bm{v})$ for each voxel $\bm{v}$ in a set of PAI volumes. For estimation of optical absorption in a voxel $\bm{v}_\textrm{u}$ of a previously unseen image, the voxel-specific CI is generated and used to infer fluence $\hat{\phi}(\bm{v}_\textrm{u})$ using the trained algorithm.\par
In our prototype implementation of the CE-qPAI method we use a random forest regressor. With voxel-based CIs, thousands of training samples can be extracted from a single slice of a simulated PA training volume. Ground truth training data generation is performed using a dedicated software plugin integrated into MITK and simulating the fluence with mitkMcxyz. It should be noted that the simulated images consist mainly of background voxels and not of vessel structures which are our regions of interest (ROI). This leads to an imbalance in the training set. To avoid poor estimation for underrepresented classes \cite{Estabrooks2004-yg}, we undersample background voxels in the training process to ensure a 1:1 ROI / background sample ratio. The parameters of the random forest are set to the defaults of sklearn 0.18 using python 2.7, except for the tree count which was set to $n_\textrm{regressors}$ = 100. CIs are used as feature vectors and labeled with the optical property to be estimated (e.g. fluence or oxygenation). The parameters were chosen based on a grid search on a separate dataset not used in the experiments of this work.\par

\subsection*{Hardware setup}
We assume a typical linear probe hardware setup \cite{Kirchner2016-rp} where the ultrasound detector array and the light source move together and the illumination geometry is the same for each image recorded. This is also the case for other typical tomographic devices \cite{Neuschmelting2016-we,Needles2013-vy}. All simulation were performed on high-end CPUs (Intel i7-5960X).

\section*{Experiments and Results}
In the following validation experiments, we quantify the fluence up to an imaging depth of 28\,mm in unseen test images for each dataset.
With our implementation and setup, all images comprise 3008 training samples, which results in an average simulation time of 2 seconds per training sample. This allows us to generate enough training samples in a feasible amount of time, to train a regressor that enables fluence estimation in a previously unseen image in near real-time. The measured computational time for quantifying fluence in a single $64\times47$ voxel image slice is 0.9\,s $\pm$ 0.1\,s.\par
In the following, we present the experimental design and results of the validation of CE-qPAI. First we will validate the estimation of absorption from PAI volumes acquired at a fixed wavelength and then estimate blood oxygenation from multispectral PAI volumes.\par

\subsection*{Monospectral absorption estimation}
\subsubsection*{Experiment}
To assess the performance of CE-qPAI in PA images of blood vessels, we designed five experimental datasets with varying complexity as listed in Table\,\ref{tabSetsIni}. With the exception of DS$_\textrm{multi}$, each of the five experimental datasets is composed of 150 training items, 25 validation items and 25 test items, where each item comprises a 3D simulated PA image of dimensions $64 \times 47 \times 62$ and 0.6\,mm equal spacing as well as a corresponding (ground truth) fluence map.\par
\begin{table}[ht!]
    \centering
    \begin{tabular}{llll}
        \toprule 
            Dataset & 
            \multirow{3}{*}{\parbox{1.5cm}{Vessel\\radius\\ $[\textrm{mm}]$}} & 
            \multirow{3}{*}{\parbox{1.9cm}{Absorption\\ coefficient\\ $\mu_\textrm{a} \,[\textrm{cm}^{-1}]$}} & 
            \multirow{2}{*}{\parbox{1.2cm}{Vessel\\count}} \\
             &  &  & \\
             &  &  & \\
            \cmidrule(lr){1-4}
            DS$_\textrm{base}$ & 3 & 4.7 & 1 \\
            \cmidrule(lr){1-4}            
            DS$_\textrm{radius}$
            &\textbf{0.5 - 6}
            &4.7
            &1\\
            
            DS$_\textrm{absorb}$
            &3
            &\textbf{1 - 12}
            &1\\
            
            DS$_\textrm{vessel}$
            &3
            &4.7
            &\textbf{1 - 7}\\
            \cmidrule(lr){1-4}            
            DS$_\textrm{multi}$
            &0.5 - 6
            &1 - 12
            &1 - 7\\

        \bottomrule
    \end{tabular}
    \caption[Dataset design.]{\textbf{Dataset design.} The design parameters of the datasets (DS)}
    \label{tabSetsIni}
\end{table}
As labels of the generated CIs we used a fluence correction $\phi_\textrm{c}(\bm{v'}) = \phi(\bm{v'}) / \phi_\textrm{h}(\bm{v'})$, where $\phi_\textrm{h}(\bm{v'})$ is a fluence simulation based on a homogeneous background tissue assumption.
We used 5 equidistant slices out of each volume, resulting in a generation of a total of 2,256,000; 376,000 and 376,000 context images for each dataset - for training, parameter optimization and testing respectively. To account for the high complexity of  DS$_\textrm{multi}$, we increased the number of training volumes for that set from 150 to 400. The baseline dataset DS$_\textrm{base}$ represents simulations of a transcutaneously scanned simplified model of a blood vessel of constant radius (3\,mm) and constant absorption (vessel: 4.73\,cm$^{-1}$, background: 0.1\,cm$^{-1}$) and reduced scattering coefficient (15\,cm$^{-1}$). To approximate partial volume effects, the absorption coefficients in the ground truth images were Gaussian blurred with a sigma of 0.6\,mm. Single slices were simulated using 10$^8$ photons and then compounded in a fully scanned volume. Different shapes and poses of the vessel were generated by a random walk defined with steps $\bm{r}$ defined as
\begin{equation}
    \bf{r}_\textrm{i} = \bf{r}_\textrm{i-1} + \eta \cdot \bf{a}   
\end{equation}
where $\eta$ is a free parameter constant in each vessel with an inter-vessel variation within a uniform distribution $( 0 < \eta < 0.2)$ and $\bm{a}$ is varied for each of its components in each step within a uniform distribution $( - 0.2\,\textrm{mm} < a_\textrm{i} < 0.2\,\textrm{mm})$. To investigate how variations in geometry and optical properties impact the performance of our method, we designed further experimental datasets in which the number of vessels (DS$_\textrm{vessel}$), the radii of the vessels (DS$_\textrm{radius}$), the optical absorption coefficients within the vessels (DS$_\textrm{absor.}$) as well as all of the above (DS$_\textrm{multi}$) were varied. We tested the robustness of CE-qPAI to this range of scenarios without retuning CI or random forest parameters.\par
While most studies assess the performance of a method in the entire image (cf. e.g. \cite{Cox2006-hw,Tarvainen2013-cc,Zemp2010-tr}), it must be pointed out that the accuracy of signal quantification is often most relevant in a defined region of interest - such as in vessels or in regions which provide a meaningful PA signal. These are typically also the regions where quantification is particularly challenging due to the strongest signals originating from boundaries with discontinuous tissue properties. To address this important aspect we validated our method, not only on the entire image, but also in the region of interest (ROI), which we define for our datasets as voxels representing a vessel and at the same time having a contrast to noise ratio (CNR) of larger than 2, to only include significant signal in the ROI. We define CNR following Walvaert and Rosseel \cite{Welvaert2013-vd} in a voxel $\bm{v}$ as\par
\begin{equation}
    \textrm{CNR} = \dfrac{S(\bm{v}) - avg(b)}{std(b)}
\end{equation}
where the $avg(b)$ and $std(b)$ are the average and standard deviation of the background signal $b$ over a simulated image slice with a background absorption coefficient of 0.1\,cm$^{-1}$ and no other structures. Using such an image without application of a noise model, we simulated an intrinsic background noise of $(4.2 \pm 2.8)$\,a.u.\par
To investigate the robustness of CE-qPAI to noise we added the following noise models to each dataset. The noise models consist of an additive Gaussian noise term applied on the signal volumes followed by a multiplicative white Gaussian noise term, similar to noise assumptions used in prior work ~\cite{Cox2006-hw,Tarvainen2013-cc}. We examined three noise levels to compare against the simulation-intrinsic noise case: 
\begin{description}
    \item[(1)]{ 2\,\% multiplicative and $(0.125 \pm 0.125)$\,a.u. additive component}
    \item[(2)]{ 10\,\% multiplicative and $(0.625 \pm 0.625)$\,a.u. additive component}
    \item[(3)]{ 20\,\% multiplicative and $(1.25 \pm 1.25)$\,a.u. additive component}
\end{description}
The additive and multiplicative noise components follow an estimation of noise components on a custom PA system~\cite{Kirchner2016-rp}. For each experimental dataset introduced in Table\,\ref{tabSetsIni} and each noise set, we applied the following validation procedure separately. Following common research practice, we used the training data subset for training of the random forest and the validation data subset to ensure the convergence of the training process, as well as to set suitable parameters for the random forest and ROI, whereas we only evaluated the test data subset to report the final results (as described in \cite{Ripley2007-qt}). As an error metric we report the \emph{relative fluence estimation error} $e_\textrm{r}$
\begin{equation}
e_\textrm{r}(\bm{v}) = \dfrac{|\hat{\phi}(\bm{v}) - \phi(\bm{v})|}{\phi(\bm{v})}
\end{equation}
rather than an absorption estimation error, to separate the error in estimating fluence with CE-qPAI from errors introduced through simulation-intrinsic or added noise on the signal which will affect the quantification regardless of fluence estimation.\par 

\subsubsection*{Results}

\begin{figure}[b!]
    \includegraphics{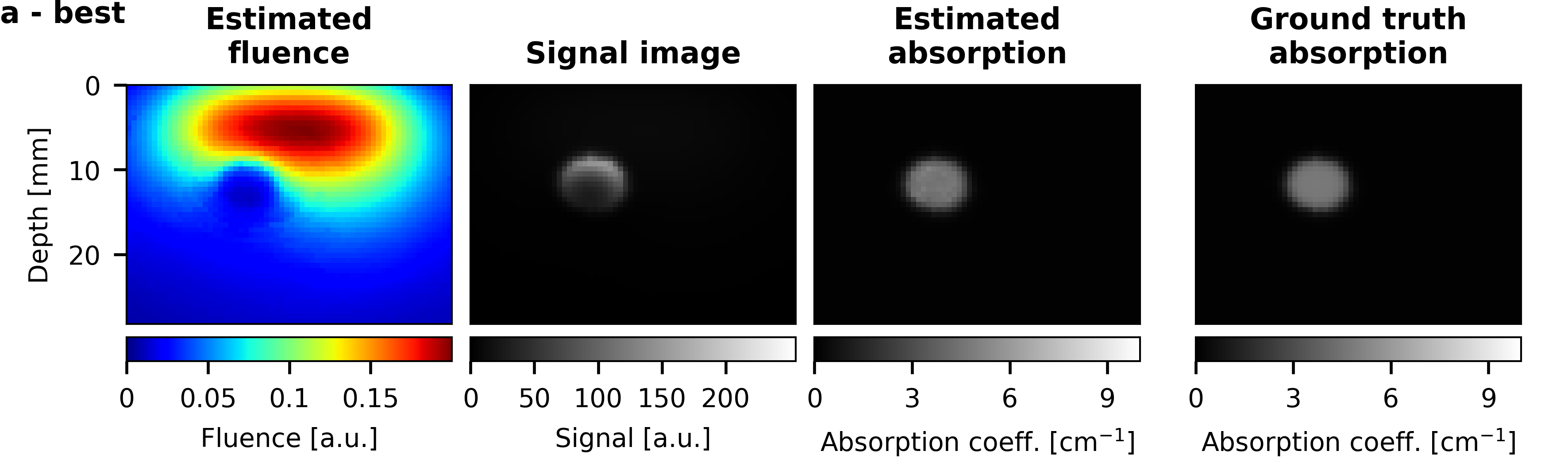}
    \includegraphics{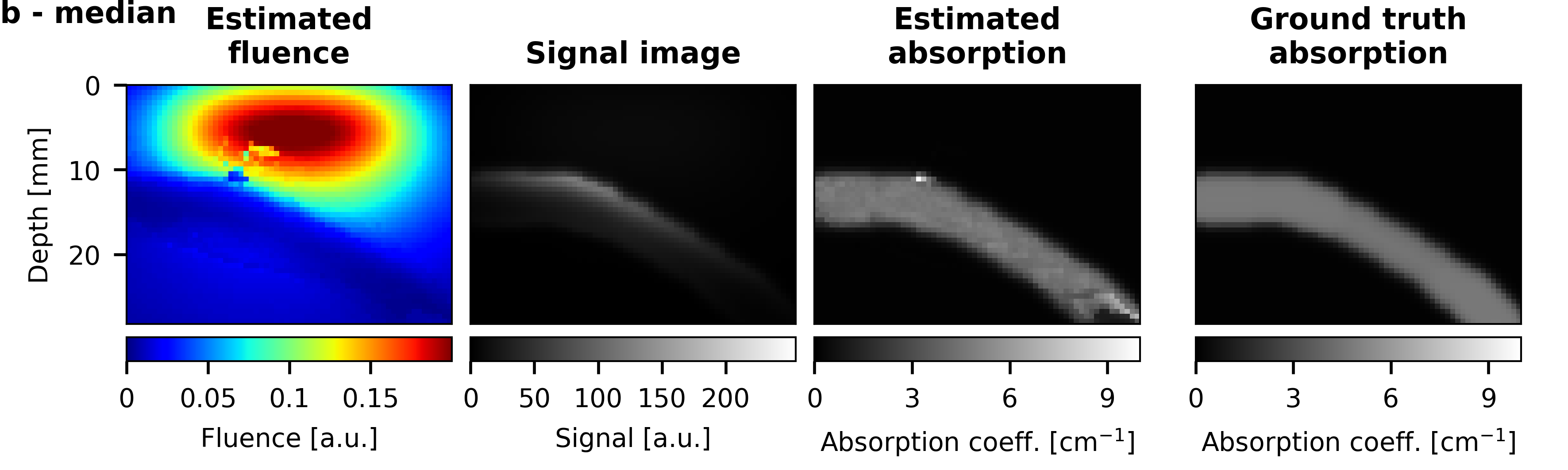}
    \includegraphics{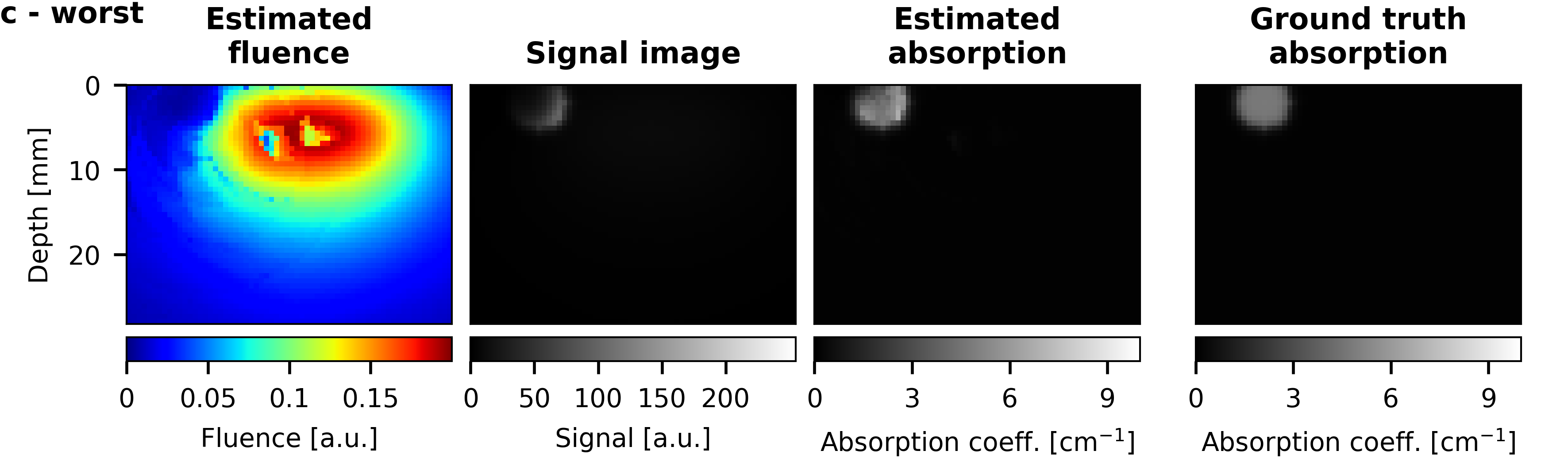}\\
    \hspace{.5cm}\\
    \includegraphics{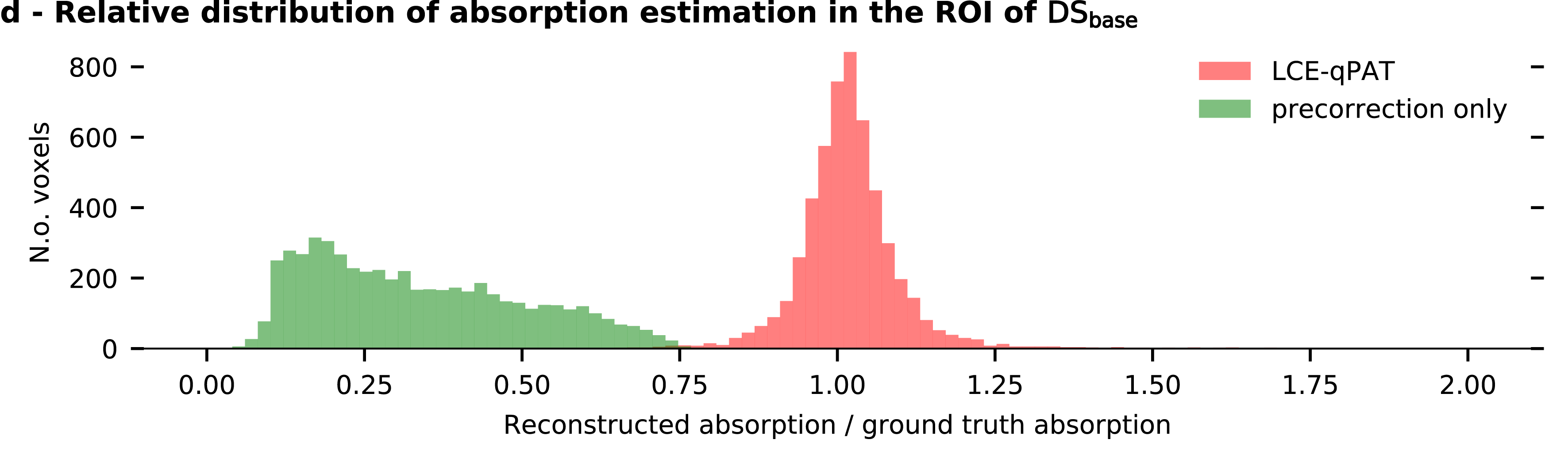}
    \caption[Absorption reconstruction results after fluence estimation]{ \textbf{Absorption reconstruction results after fluence estimation.} For the slices with the \textbf{a} lowest, \textbf{b} median, and \textbf{c} highest median fluence estimation error $e_\textrm{r}$ within the region of interest (ROI) of DS$_\textrm{base}$. We show (from left to right) (1) the estimated fluence, (2) the corresponding signal images, (3) the resulting estimation of the absorption coefficient, and (4) the ground truth optical absorption, for reference. \textbf{d} shows a histogram of the relative absorption estimation over all ROI voxels ($n=5347$) in DS$_\textrm{base}$ illustrating absorption estimation accuracy rather than fluence estimation accuracy measured by $e_\textrm{r}$. Precorrecting the signal with the fluence of a homogeneous tissue assumption underestimates the absorption and is considerably outperformed by CE-qPAI in the ROI. The CE-qPAI plot omits 5 outliers larger 2.}
    \label{figReconstruction}
\end{figure}

Figure\,\ref{figReconstruction}\,a-c show representative examples of the previously unseen 125 simulated test images from the baseline dataset DS$_\textrm{base}$, with their corresponding fluence estimation results. The optical absorption is reconstructed using the fluence estimation. A histogram illustrating absorption estimation accuracy in region of interest (ROI) voxels of DS$_\textrm{base}$ is shown in Figure\,\ref{figReconstruction}\,d and compared with a static fluence correction approach.

\begin{table}[htb!]
    \centering
    \begin{tabular}{lrrrr}
        \toprule 
            & \multicolumn{4}{c}{Relative error $e_\textrm{r}$}\\
            \cmidrule(lr){2-5}
            & \multicolumn{2}{c}{All voxels} & \multicolumn{2}{c}{ROI}\\
            \cmidrule(lr){2-3}
            \cmidrule(lr){4-5}
            Dataset & Median & \multicolumn{1}{c}{IQR} & Median & \multicolumn{1}{c}{IQR} \\
            \cmidrule(lr){1-5}
            DS$_\textrm{base}$ & 0.5\,\% & (0.2, 1.0)\,\% &3.9\,\% &(1.9, 7.3)\,\% \\
            
            DS$_\textrm{radius}$ &0.8\,\%
            &(0.3, 2.7)\,\%
            &5.8\,\%
            &(2.6, 11.4)\,\% \\
            
            DS$_\textrm{absorb}$ &0.7\,\%
            &(0.3, 1.9)\,\%
            &14.0\,\%
            &(5.0, 31.5)\,\% \\
            
            DS$_\textrm{vessel}$ &1.6\,\%
            &(0.5, 5.9)\,\%
            &6.8\,\%
            &(2.9, 13.4)\,\% \\

            DS$_\textrm{multi}$ &1.2\,\%
            &(0.4, 5.4)\,\%
            &20.1\,\%
            &(7.3, 49.0)\,\% \\

        \bottomrule
    \end{tabular}
    \caption[Descriptive statistics of fluence estimation results.]{\textbf{Descriptive statistics of fluence estimation results.} The median and interquartile range (IQR) of the relative fluence estimation error $e_\textrm{r}$ for the five validation datasets used for the single wavelength experiments. The median error and IQR are provided (1) for all voxels in the respective test set as well as (2) for the voxels in the region of interest (ROI) only.}
    \label{tabSets}
\end{table}
\begin{figure}[htb!]
    \includegraphics{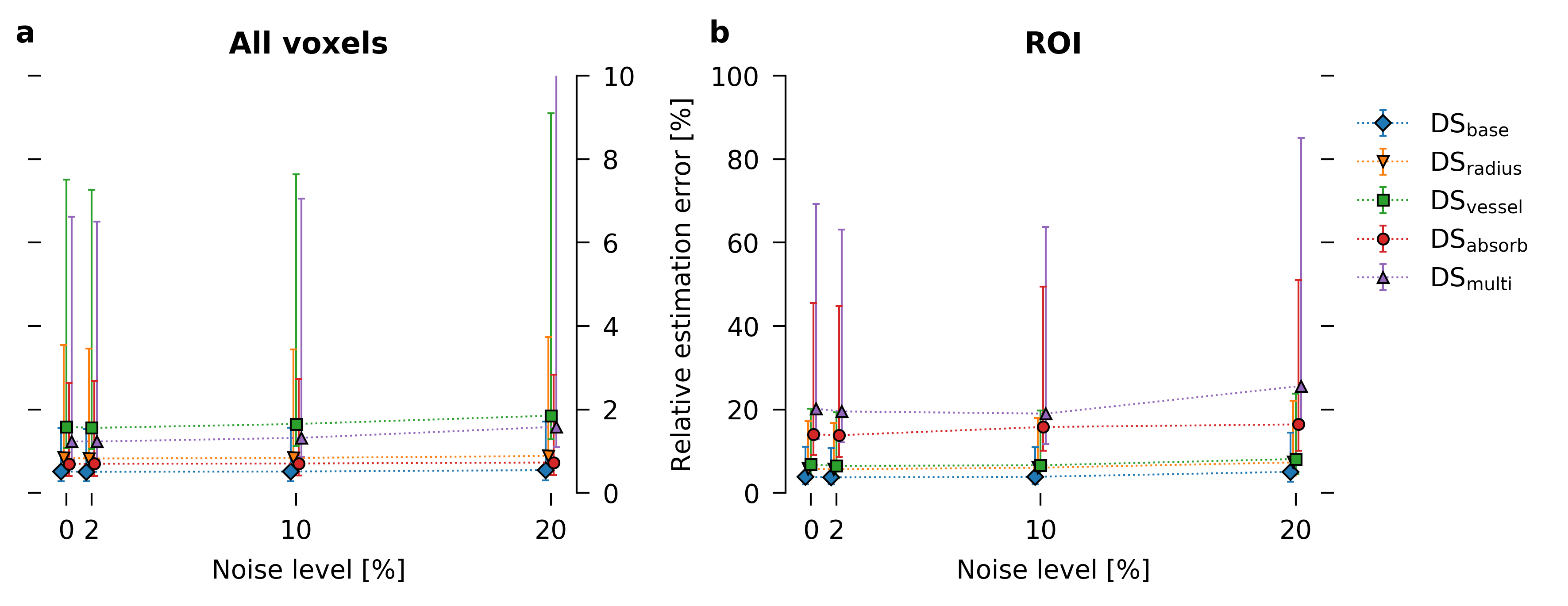}
    \caption[Robustness of the fluence estimation against noise.]{\textbf{Robustness of the fluence estimation against noise.} Median relative fluence estimation errors $e_\textrm{r}$ with interquartile range over all datasets for, \textbf{a} all test voxels, and \textbf{b} in region of interest test voxels. The whiskers in this plot show the first and third quartile.}
    \label{figRobustness}
\end{figure}

Table\,\ref{tabSets} summarizes the descriptive statistics of the relative fluence estimation errors $e_\textrm{r}$ for the experiments on absorption estimation using single wavelength PA images. The relative fluence estimation error $e_\textrm{r}$ does not follow a normal distribution due to large outliers especially in complex datasets, which is why we report median $e_\textrm{r}$ with interquartile ranges (IQR) for all datasets. Even for the most complex dataset DS$_\textrm{multi}$ with variations of multiple parameters, specifically, number of vessels, vessel absorption coefficient and vessel radii CE-qPAI yields a median overall relative fluence estimation error $e_\textrm{r}$ below 2\,\%. Errors are higher in the ROI, especially in datasets with high variations of absorption.\par

Previously proposed qPAI approaches reveal high drops in estimation performance when dealing with noisy data (cf. e.g. \cite{Beretta2016-qc}). To remedy this, methods have been proposed to incorporate more accurate noise representations into model based reconstruction algorithms \cite{Tarvainen2013-cc,Tarvainen2016-yq}. When validating the robustness of CE-qPAI to noise, it yields high accuracy even under unrealistically high noise levels of up to 20\,\% (cf. Figure\,\ref{figRobustness}). Regardless of the noise level applied, the highest median errors occur in the ROIs of datasets which are characterized by high absorption and inhomogeneous tissue properties.\par

\subsection*{Multispectral blood oxygenation estimation}
The concept of context encoding cannot only be used to estimate fluence and absorption but also derived functional parameters such as blood oxygenation. To this end, the estimated absorption in a voxel for multiple wavelengths can be applied to resolve oxygenation via linear spectral unmixing. Alternatively, a regressor can be trained using the CIs labeled with ground truth oxygenation.\par 
\subsubsection*{Experiment}
To investigate the performance of CE-qPAI for blood oxygenation (sO$_2$) estimation we designed an additional multispectral simulated dataset  DS$_\textrm{oxy}$ using the wavelengths 750\,nm, 800\,nm and 850\,nm. It consists of 240 multispectral training volumes and 11 multispectral test volumes, each featuring homogeneous oxygenation and one vessel with a radius of 2.3 to 4\,mm - modeled after a carotid artery~\cite{Krejza2006-wj}. For each image slice and at each wavelength, $10^7$ photons were used for simulation. Oxygenation values for the training images were drawn randomly from a uniform sO$_2$ distribution $U(0\,\%, 100\,\%)$. For testing, we simulated 11 multispectral volumes at 3 wavelengths and 11 blood oxygenation levels  (sO$_2 \in \{ 0\,\%, 10\,\%, 20\,\%, \dots , 100\,\% \}$). The optical absorption was adjusted by wavelength and oxygenation, as described by Jacques~\cite{Jacques2013-pm}. Hemoglobin concentration was assumed to be 150\,g/liter~\cite{Jacques2013-pm}. The blood volume fraction was set to 0.5\,\% in the background tissue and to 100\,\% in the blood vessels. The reduced scattering coefficient was again set to 15\,cm$^{-1}$. 
We estimated the oxygenation using three methods:\par
(1) \emph{Linear spectral unmixing on the signal images as a baseline \cite{Keshava2002-wv}.} For this, we applied a non-negative constrained least squares approach as also used in \cite{Tzoumas2016-yt} that minimizes $||A\bm{x}-\bm{b}|| = 0$, where $A$ is the matrix containing the reference spectra, $\bm{b}$ is the measurement vector, and $\bm{x}$ is the unmixing result. Specifically, we used the python scipy.optimize.minimize function with the Sequential Least SQuares Programming (SLSQP) method and added a non-negativity inequality constraint. We evaluated the unmixing results of this method on all voxels in the ROI as well as exclusively on those voxels with the maximum intensity projection (MIP) along image x-axis at wavelength 800\,nm to account for nonlinear fluence effects deep inside the vessels.\par
(2) \emph{Linear spectral unmixing of the signal after quantification of the three input images with CE-qPAI.} After correcting the raw signal images for nonlinear fluence effects using CE-qPAI, we applied the same method as described in (1) and evaluated on the same voxels that were used in (1) to ensure comparability of the results.\par
(3) \emph{Direct estimation of oxygenation using a functional adaptation of CE-qPAI.} For functional CE-qPAI (fCE-qPAI), triples of CIs for the three chosen wavelengths were concatenated into one feature vector and labeled with the ground truth oxygenation.

\subsubsection*{Results}
\begin{figure}[b!]
    \includegraphics{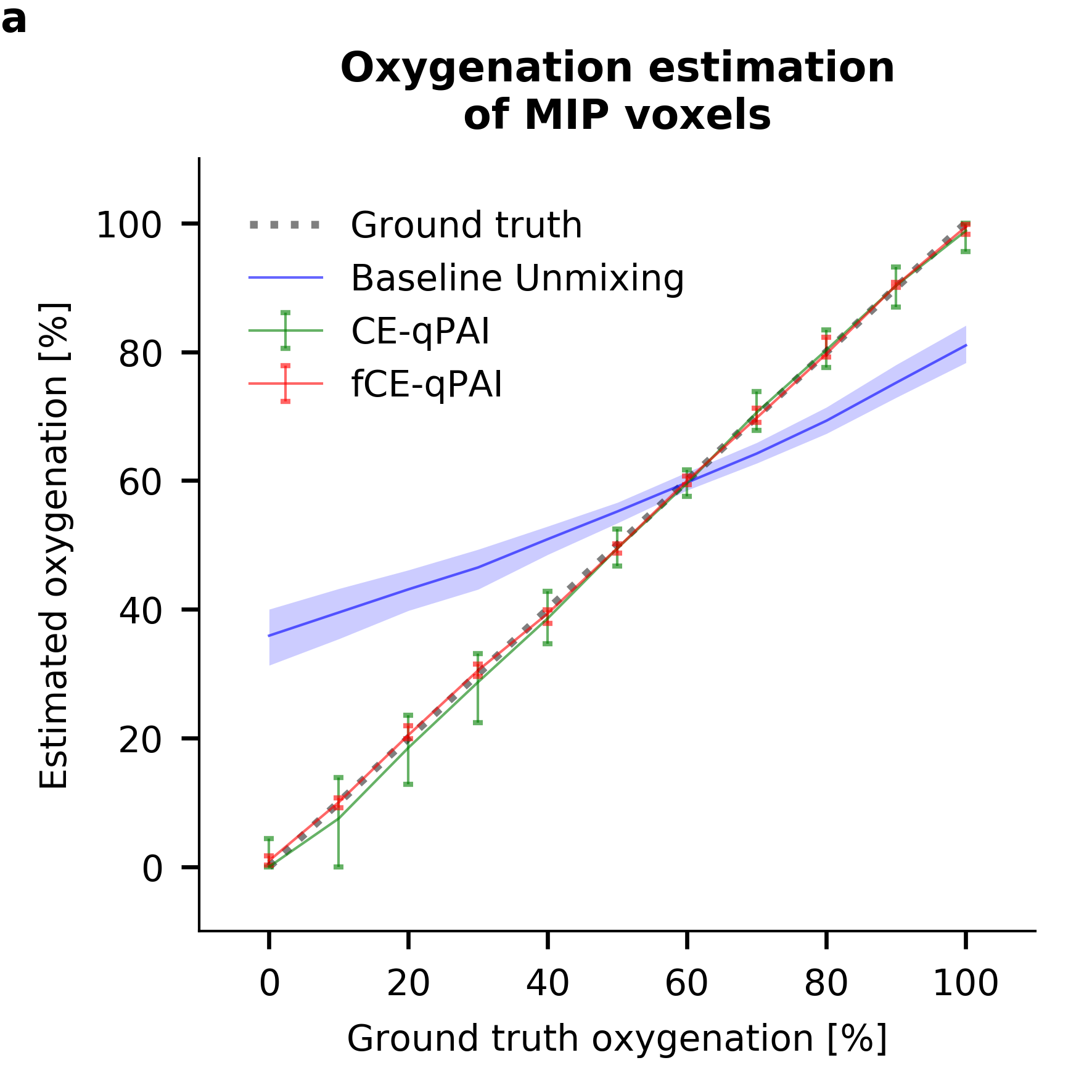}
    \includegraphics{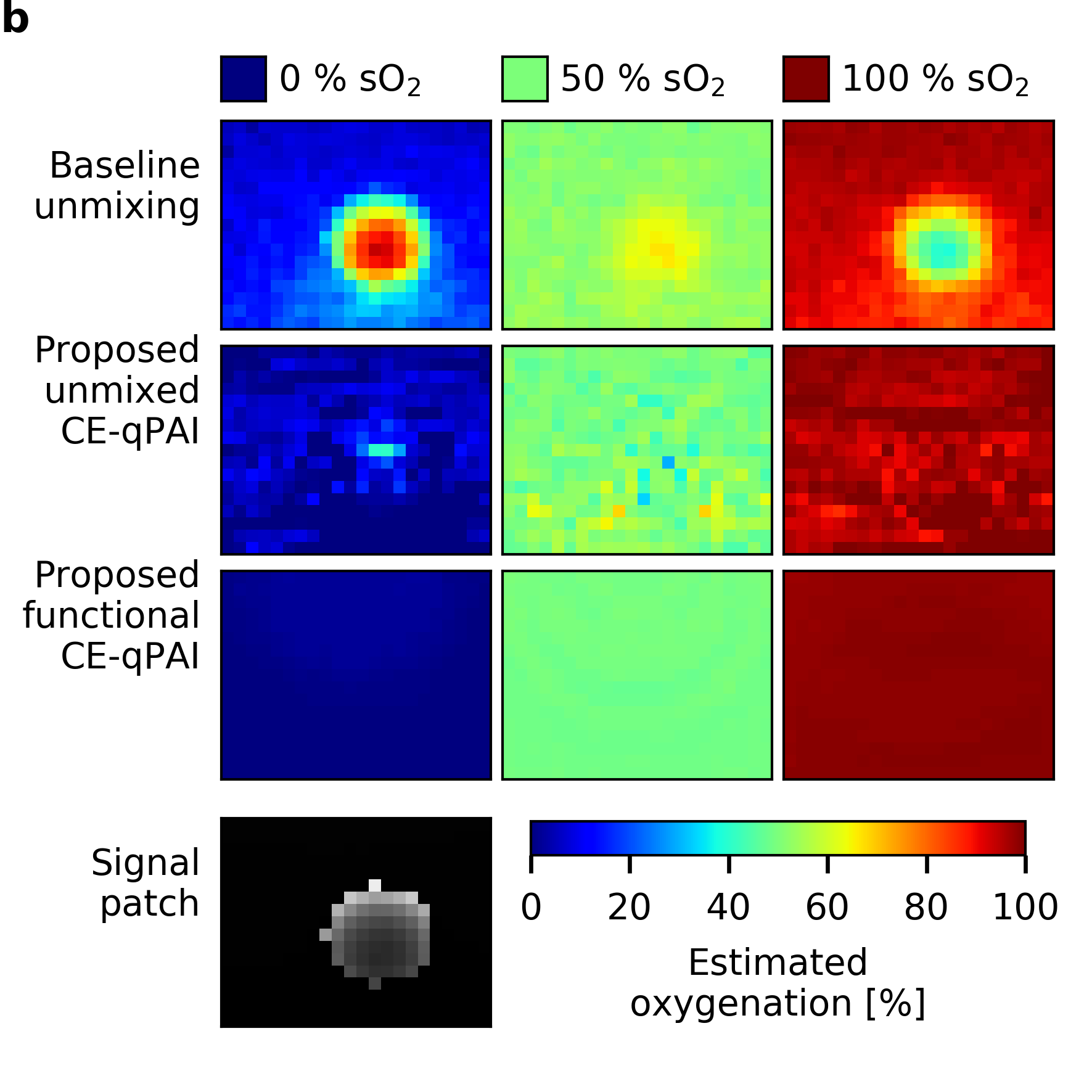}
    \caption[Performance of oxygenation estimation.]{\textbf{Oxygenation estimation.} \textbf{a} shows the median oxygen estimation with the interquartile range (IQR) on the maximum intensity projection (MIP) voxels using linear spectral unmixing of (blue) the uncorrected signal, (green) the signal corrected by CE-qPAI, and (red) direct estimation by functional CE-qPAI (fCE-qPAI). \textbf{b} shows the oxygenation estimation for a representative patch of signal showing a vessel in 15\,mm depth and with 3\,mm radius. The signal for one of the measurement wavelengths is shown for reference together with the oxygen estimation results for 0\,\%, 50\,\%, and 100\,\% ground truth homogeneous oxygenation and the three examined methods.}
    \label{figOxygenation}
\end{figure}
Estimation of local blood oxygen saturation (sO$_2$) is one of the main qPAI applications and is only possible with multispectral measurements. As such, the presented approaches were validated together with the baseline method on the dataset DS$_\textrm{oxy}$. As shown in Figure\,\ref{figOxygenation}a, the estimation results for both methods are in very close agreement with the ground truth. In fact, the median absolute oxygen estimation error was 3.1\,\% with IQR (1.1\,\%, 6.4\,\%) for CE-qPAI and 0.8\,\% with IQR (0.3\,\%, 1.8\,\%) for the fCE-qPAI adaptation. Furthermore, our methodology outperforms a baseline approach based on linear spectral unmixing of the raw signal (as also compared to in \cite{Tzoumas2016-yt}). By means of example Figure\,\ref{figOxygenation}b shows that linear spectral unmixing of the ROI on the uncorrected signal fails deep inside the ROI where the fluence varies strongly for different wavelengths. To compensate for this effect when comparing the approach to our method, we validate all methods only on the maximum intensity projection along the depth axis (as also used in \cite{Dean-Ben2014-uu}) in Figure\,\ref{figOxygenation}a.\par

\section*{Discussion}
This paper addresses one of the most important challenges related to photoacoustic imaging, namely the quantification of optical absorption based on the measured signal. In contrast to all other approaches proposed to qPAI to date (cf. e.g. \cite{Cox2009-vn,Iftimia2000-bb,Cox2005-tl,Cox2006-hw,Yuan2006-qx,Laufer2007-va,Malone2016-ew,Haltmeier2015-eq,Cox2012-ao,Banerjee2008-rj}), our method relies on \textit{learning} the light fluence in a voxel to deduce the corresponding optical absorption. Comprehensive \emph{in silico} experiments presented in this manuscript show the high potential of this entirely novel approach to estimate optical absorption as well as derived functional properties, such as oxygenation, even in the presence of high noise.\par
Although machine learning methods have recently been applied to PAI related problems (cf. e.g. \cite{reiter2017-ml,hauptmann2017-mb,antholzer2017-dl}), these have mainly focused on image reconstruction but not signal quantification. We attribute this to the fact that training generation for machine learning based qPAI is not at all straightforward given the lack of reference methods for estimating optical absorption in depth and the long simulation times of Monte Carlo based methods. Note also that commonly applied methods of data augmentation (i.e. methods that may be used to automatically enlarge training data sets as discussed in \cite{Dosovitskiy2014}) cannot be applied to PA images due to the interdependence of fluence and signal. There have been recent developments, however, that could speed up training data generation using hybrid diffusion approximation and Monte Carlo methods \cite{zhu2012-hm}.
With our contribution, we have addressed the challenge by introducing the concept of context images, which allow us to generate one training case from each \textit{voxel} rather than from each image. \par
As an important contribution with high potential impact, we adapted CE-qPAI to estimate functional tissue properties from multi wavelength data. Both variants - linear spectral unmixing of the fluence corrected signal, as well as direct estimation of oxygenation from multi wavelength CIs, yielded accurate results that outperformed a baseline approach based on linear spectral unmixing of the raw PA signal. It should be noted that linear spectral unmixing of the signal for sO$_2$ estimation is usually performed on a wider range of wavelengths to increase accuracy. However, even this increase in number of wavelengths cannot fully account for nonlinear fluence effects \cite{Cox2009-vn}. Combined with the separately established robustness to noise, multi wavelength applications of CE-qPAI are very promising.\par
In our first prototype implementation of CE-qPAI we used random forests regressors with standard parameters. It should be noted, however, that fluence estimation from the proposed CI can in principle be performed by any other machine learning method in a straightforward manner. Given the recent breakthrough successes of convolutional neural networks \cite{He2016-vv}, we expect even better performance of our approach when applying deep learning algorithms or using data augmentation in the training data generation process.\par
By relating the measured signals $\textrm{S}(\bm{v'})$ in the neighborhood of $\bm{v}$ to the corresponding fluence contributions $\textrm{FCM}[\bm{v}](\bm{v'})$ we relate the absorbed energy in $\bm{v'}$, to the fluence contribution of $\bm{v'}$ to $\bm{v}$. In this context it has to be noted that the fluence contribution $\textrm{FCM}[\bm{v}](\bm{v'})$ is only an approximation of the true likelihood that a photon passing $\bm{v}$ has previously passed $\bm{v'}$, because $\textrm{FCM}[\bm{v}]$ is generated independently of the scene under observation assuming constant background absorption and scattering. Nevertheless due to the generally low variance of scattering in tissue it serves as a reliable input for the proposed machine learning based quantification.\par
A limitation of our study can be seen in the fact that we performed the validation \textit{in silico}. To apply CE-qPAI \emph{in vivo}, further research will have to be conducted in two main areas. Firstly, the acoustical inverse problem for specific scanners must be integrated into the quantification algorithm to enable quantification of images acquired with common PAI probes such as clinical linear transducers.  Secondly, training data has to be generated as close to reality as possible - considering for example imaging artifacts.\par
In contrast to prior work (cf. e.g.  \cite{Cox2006-hw,Yuan2006-qx,Zemp2010-tr,Tarvainen2013-cc,Naetar2014-zu}) our initial validation handles the whole range of near infrared absorption in whole blood at physiological hemoglobin concentrations and demonstrates high robustness to noise. The impact of variations of scattering still needs investigation although these should be small in the near infrared.\par
Long-term goal of our work is the transfer of CE-qPAI to \emph{clinical} data. In this context, run-time of the algorithm will play an important role. While our current implementation can estimate absorption on single slices within a second, this might not be sufficient for interventional clinical estimation of whole tissue volumes and at higher resolutions. An efficient GPU implementation of the time intensive CI generation should enable real-time quantification.\par
In summary, CE-qPAI is the first machine learning based approach to quantification of PA signals. The results of this work suggest that quantitative real-time functional PA imaging deep inside tissue is feasible.

\section*{Disclosures}
The authors have no relevant financial interests in this article and no potential conflicts of interest to disclose.

\section*{Acknowledgements}
The authors would like to acknowledge support from the European Union through the ERC starting grant COMBIOSCOPY under the New Horizon Framework Programme grant agreement ERC-2015-StG-37960.\par
We would like to thank the ITCF of the DKFZ for the provision of their computing cluster, C.Feldmann for her support with figure design. A.M.Franz, A.Seitel, F.Sattler, S.Wirkert and A.Vemuri for reading the manuscript.

\section*{Author Contributions}
T.K., J.G. and L.M. conceived of the research, analyzed the results and wrote the manuscript. T.K. and J.G. wrote the software and performed the experiments. L.M. supervised the project.

\section*{Code and Data Availability}
The code for the method as well as the experiments was written in C++ and python 2.7 and is partially open source and available at https://phabricator.mitk.org/source/mitk.git. Additional code and all raw and processed data generated in this work is available from the corresponding authors on reasonable request.

\end{document}